# Free Radicals in Superfluid Liquid Helium Nanodroplets: A Pyrolysis Source for the Production of Propargyl Radical

Jochen Küpper, Jeremy M. Merritt, and Roger E. Miller
University of North Carolina,
Department of Chemistry,
Chapel Hill, NC 27599, USA

**Abstract**

An effusive pyrolysis source is described for generating a continuous beam of radicals under conditions appropriate for the helium droplet pick-up method. Rotationally resolved spectra are reported for the $\nu_1$ vibrational mode of the propargyl radical in helium droplets at 3322.15 cm$^{-1}$. Stark spectra are also recorded that allow for the first experimental determination of the permanent electric dipole moment of propargyl, namely -0.150 D and -0.148 D for ground and excited state, respectively, in good agreement with previously reported *ab initio* results of -0.14 D [1]. The infrared spectrum of the $\nu_1$ mode of propargyl-bromide is also reported. The future application of these methods for the production of novel radical clusters is discussed.

**Introduction**

Although free radicals play a central role in many gas phase chemical processes, including combustion, their spectroscopic study is often hampered by their reactivity, which limits the gas phase number densities that can be achieved in practical experiments. In recent years, however, a number of methods have been developed for producing such radicals in sufficiently high concentrations to permit high-resolution spectroscopic studies in the gas phase [2-24]. In the infrared region of the spectrum, direct absorption methods have been used to obtain rotationally resolved spectra of a number of such systems, in both gas cells [3,4] and free jet expansions [5-16]. The radicals of interest are typically produced by pyrolysis [5,6], microwave discharge [7-12], electric discharge [13-16] or photolysis [17-24]. For example, flash pyrolysis has been used to generate radicals in cold supersonic jets, as first demonstrated by Kohn et al. [5]. In these and subsequent studies, small organic radicals were produced from the pyrolysis of halogenated precursors seeded in helium or argon. It was often possible to obtain essentially complete depletion of the precursor molecules [25].

In light of its role in sooting flames, propargyl (2-propynyl) radical (Structure I) has been the focus of particular attention [26-29]. Propargyl is one of the simplest conjugated systems with an odd number of electrons, also making the focus of considerable theoretical study [1,30-32]. There is now compelling evidence that propargyl is the most important radical precursor in the formation of benzene, polycyclic aromatic hydrocarbons (PAH), and soot in certain combustion processes [26-29,33-38]. For example, the simple dimerization of two propargyl radicals is thought to be important in the formation of benzene, as suggested by Wu and Kern [33].

Early observations of the propargyl radicals where carried out by ESR spectroscopy in liquid allene [39] and in argon matrix studies [40]. Ramsey and Thistlethwaite [41] first observed the electronic gas-phase absorption spectrum. Infrared spectra and electronic absorption spectra were obtained in neon and argon matrices [42-44], including the study by Jacox and Milligan [42], which identified four vibrational bands (at 3308, 686, 548 and 484 cm$^{-1}$)

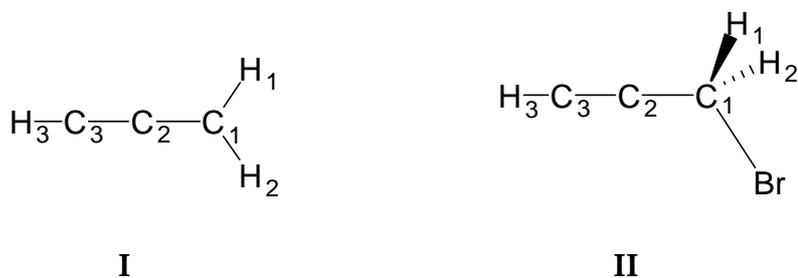

**I**                  **II**

that they assigned to propargyl. Later studies by Maier et al. [45], confirmed by Huang and Graham [44], showed that the 548 cm$^{-1}$ band is actually associated with triplet propargylene. Recently the infrared and electronic absorption spectrum of propargyl in neon matrices has been examined [43].

Additional information on the propargyl radical has been obtained from the photoelectron spectrum of the allenyl anion [46], i.e. a long progression with a mean spacing of approximately 515 cm$^{-1}$ was assigned to an out-of-plane bending motion.

Propargyl radical has also been produced in supersonic jets by means of flash pyrolysis, for use in photoelectron studies [47-49]. Rotationally resolved spectra of propargyl were first obtained for the $\nu_1$ acetylenic CH-stretching vibration at 3322.3 cm$^{-1}$ by the group of Curl [3]. This work was extended to the $\nu_6$ vibrational band by Tanaka et al. [50] and to higher rotational levels of the $\nu_1$ band [4]. In addition, Fourier Transform microwave spectroscopy has been reported for this radical [51]. All of these rotationally resolved studies confirmed the $C_{2v}$ symmetry of the electronic ground state of propargyl [3,4,50,51], in agreement with ab initio quantum mechanical calculations [1,30-32,52], most recently at the coupled cluster level by Botschwina et al. [30].

In light of the importance of propargyl radicals in combustion, we have started a program of study aimed at investigating the associated interactions in the pre-reactive dimer and between propargyl and other molecules and atoms. The approach taken is to solvate the radicals in helium droplets, which act as a nearly ideal matrix for spectroscopic study [53-60]. In the present study we focus on the spectroscopy of the radical monomer. As part of this research program we report here the development and use of a novel pyrolysis source designed to produce a clean beam of radicals at the low fluxes needed for helium droplet pick-up experiments. In the present study we pyrolyse propargyl bromide (Structure II) to produce the radical. Rotationally resolved infrared spectra of the radical are obtained using the helium droplet apparatus already developed in our laboratory for this purpose [61]. The source described herein has also been successfully applied to the generation of Br atoms and CN radicals [62].

**Experimental**

The experimental apparatus used in the present study has been described in detail previously [58]. Therefore the present discussion focuses on the pyrolysis source and only briefly reviews the overall apparatus. The pyrolysis-source is shown schematically in Figure 1. It consists of a 72 cm long glass tube with an outer diameter of 6 mm. The gas of interest flows through a 64 cm section of pyrex tubing, followed by a graded seal junction of 3 cm and a 5 cm long quartz tip. The quartz tip is wrapped with either 0.25 mm tantalum wire or tantalum/tungsten (90/10) ribbon for resistance heating. In the former case the wire is coated with ceramic paste (Aremco Ceramacast 576/Aremco Ceramabond 552) and a K-type thermocouple is embedded in the ceramic for temperature measurement. A water-cooled copper block is used to cool the glass tube approximately 7.5 cm from the tip.

The tip can be operated routinely at 1400 K and up to 1800 K for short periods of time. The flow of propargyl bromide, from a static reservoir at room-temperature, is regulated by a stainless steel needle valve. The total pressure in the pyrolysis region is estimated to be 3·10$^{-4}$ mbar. At these low pressures the molecules undergo only a few collisions with the walls and essentially no collisions in the gas phase. This helps to minimize recombination of the newly formed radicals. At temperatures above 1000 K the propargyl bromide spectrum can no longer be detected, due to the complete dissociation of the precursor.

The overall experimental setup is housed in two differentially pumped vacuum chambers. In the first chamber helium gas is expanded from 50–90 bar into vacuum through a 5 $\mu$m orifice cooled to 20–24 K. This results in the formation of droplets with a mean size of approximately 3000–7500 helium atoms. After passing through a skimmer into the second chamber the droplet beam passes approximately 1 mm in front of the exit from the effusive pyrolysis source, as shown in Figure 1. The transparency of the droplets to black body radiation is illustrated by the fact that they pass so close to a 1400 K surface without being evaporated. The doped droplets then pass through the infrared beam from a color center laser (Burleigh FCL-20), reflected between two parallel gold-coated mirrors. In the present study, the laser was operated on crystal #3 (RbCl:Li), providing continuously tunable radiation from 3000–3500 cm$^{-1}$. Vibrational excitation of the $\nu_1$ stretch of the radical is followed by relaxation of the associated energy into the helium droplet. Evaporation of helium atoms from the droplet results in a reduction of the helium flux to the bolometer detector. The laser beam is chopped and the signals are processed by a phase-sensitive detector. The laser interaction region lies between two electrodes, used to apply electric fields of up to 100 kV/cm. In the present study the applied fields are used to measure the dipole moment of the propargyl radical.

**Results and Discussion**

As noted above, the propargyl radicals were generated in this study by pyrolysis of propargyl bromide. To aid in the optimization of the radical source we first monitored the helium solvated propargyl bromide by exciting the corresponding $\nu_1$ (C-H stretching) vibration. Propargyl bromide has been considered a replacement for methyl bromide in soil fumigation [63]. Its basic physical and environmental properties [63,64] and its infrared spectrum [65,66] have been studied previously. The helium droplet spectrum is shown in Figure 2. To aid in the analysis of this spectrum we performed *ab initio* calculations at the Hartree-Fock, MP2, and CCSD(T) levels. All calculations were performed in the cc-pVTZ basis [67] using MOLPRO [68]. The $C_s$ symmetry of propargyl bromide was applied and the ten remaining geometric parameters were optimized. The resulting geometries and rotational constants are given in Table 1. From the normal coordinate analysis, the transition moment orientation for the $\nu_1$ vibration is determined to be in the *ab* plane, at an angle of $\theta=40°$ with respect to the *a* axis. Given that the propargyl bromide spectrum is only partially resolved, the fit was carried out using a symmetric top Hamiltonian with the upper and lower state constants constrained to be equal.  The calculation shown in Figure 2 corresponds to $A'=A''=0.026\,\text{cm}^{-1}$, $B'=C'=0.0148\,\text{cm}^{-1}$, $B''=C''=0.0152\,\text{cm}^{-1}$, $\nu_0=3332.56$, and a linewidth of $0.033\,\text{cm}^{-1}$.  Based on the *ab initio* calculations the *B* and *C* rotational constants in helium are approximately 4.5 times smaller than those of the molecule in vacuum.  This is consistent with results obtained previously for closed shell molecules [69].  The more surprising result is that the *A* constants are approximately 20 times smaller than the *ab initio* value.  Although this anomalously large reduction in the A rotational constant might be due to an unusually strong interaction between the molecule and the helium matrix, resulting from the highly polarizeable bromine atom, at this point we do not have a good explanation for this effect.  It is worth noting, however, that the value of A is rather sensitive to the relative contributions from the *a*- and *b*-type components of the hybrid band, which in the present calculation was simply fixed at the *ab initio* value.  The experimental vibrational origin ($3332.56\,\text{cm}^{-1}$) is consistent with the low-resolution gas-phase results ($3335\pm6\,\text{cm}^{-1}$) [65,66].

The results presented above provide unambiguous identification of the propargyl bromide in the droplets. The spectrum was observed to diminish in intensity and finally disappear as the temperature of the pyrolysis source was increased. Once these conditions were established, a search was carried out for the propargyl radical, using the gas phase results as a guide.  Figure 3 shows the resulting spectrum, which can be assigned to the $\nu_1$ vibrational band of propargyl radical embedded in superfluid liquid helium droplets. Table 2 gives a list of the transition frequencies and rotational assignments for the observed transitions.  This spectrum corresponds to an *a*-type band of a prolate asymmetric top.  Transitions are observed that originate from $K_a=1$ levels even though we would not expect these states to be populated based on the temperature of the droplets (0.37 K). These results confirm that this radical has $C_{2v}$ symmetry even when embedded in liquid helium droplets, such that the $K_a=1$ levels cannot cool to $K_a=0$.  Since this spectrum is *a*-type, and the relative populations of the $K_a=1$ and $K_a=0$ states are not controlled thermodynamically the $A''$ rotational constant cannot be determined from the data.  Within the experimental uncertainty, $\Delta A$ was also the same as in the gas phase.  Therefore, for the purpose of fitting the spectra, $A''$ and $A'$ were fixed at their gas-phase values, namely $9.61\,\text{cm}^{-1}$ and $9.60\,\text{cm}^{-1}$, respectively [4,50,51].  This is a reasonable approximation given that previous studies [69] have shown that the helium cannot follow such fast rotational motion, so that the rotational constants are approximately the same as in the gas phase [70]. The five $K_a=0$ lines given in Table 2 were fit to a linear-rotor Hamiltonian, yielding $(B''+C'')/2$, $(B'+C')/2$, $D_J''$, and $D_J'$.  The values of *B-C* were determined for both vibrational states by simultaneously fitting the field-free and Stark-spectrum (Figure 4).   The resulting molecular parameters are reported in Table 3.  The $(B+C)/2$ constants in both vibrational states are reduced by a factor of 2.6 compared to the gas-phase values [4,50,51], in good agreement with previous studies of closed shell molecules in helium droplets [69,71].  The centrifugal distortion constants are much larger than for the isolated gas-phase molecule [4,50,51]. These constants are presumably more indicative of the strength of the propargyl-helium interactions than of the inherent rotation-vibration coupling in propargyl [72].

High-resolution gas-phase infrared spectra have revealed line splittings due to spin-rotation coupling (Hund's case (*b*)) [50,51]. The spin-rotation coupling constants determined by Tanaka et al. from the $v_6$ band [50] and the microwave spectrum [51] are small. In fact, Yuan et al. [4] were unable to determine spin-rotation coupling parameters even from the gas phase spectrum of the $v_1$ band. At the resolution of the present helium droplet spectrum, where only the lowest *J* and $K_a$ states of the molecule are observed, we would not expect to resolve splittings due to spin-rotation coupling. This accounts for why we are able to explore the entire spectrum using a conventional asymmetric top Hamiltonian. Although the source of the line broadening in the helium droplet spectrum is not fully understood, there is considerable evidence from previous studies that inhomogeneous effects are significant and that these are dependent upon the rotational state of the system. The best fits of the spectra were obtained with two different linewidths for the $K_a$=0 and $K_a$=1 states, namely 0.0225 cm$^{-1}$ (0.0175 cm$^{-1}$) and 0.045 cm$^{-1}$ (0.035 cm$^{-1}$) for the field free (Stark) spectrum, respectively. Since some of the $K_a$=1 transitions correspond to upper states that are the lowest rotational levels of their respective spin-symmetry, the excessive broadening here cannot be attributed to reduced rotational lifetimes. One possibility is that the spin-rotation coupling is larger for the $K_a$=1 levels, in agreement with the fact that spin-rotation-coupling is (at least) approximately linear with the overall angular momentum apart from spin.

The manner in which the intensities in the calculated spectrum were determined deserves some attention. Given the $C_{2v}$ symmetry of propargyl care must be taken to properly account for the nuclear spin statistics. The ground electronic state of propargyl has $^2B_1$ symmetry, giving rise to spin statistical weights of 3:1 for $K_a$ even and odd levels, respectively. At the temperature the propargyl radicals are formed in the radical source (approx. 1000 K) the total population in all of the $K_a$ even states will be three times the sum of population in the $K_a$ odd states. The corresponding ground state populations in the helium droplets can then be obtained by cooling all the $K_a$ even states into $K_a$=0 and the $K_a$ odd states into $K_a$=1, assuming no nuclear spin conversion. This preserves the ratio of 3:1 for the $K_a$ even and odd states. For a given $K_a$ the relative populations of corresponding rotational states are computed using a Boltzmann formula, the zero in energy corresponding to the lowest energy state within each $K_a$ manifold. The fact that the calculated intensities are in excellent agreement with the experimental results suggests that on the timescale of the present experiment (~1 ms) there is essentially no spin-relaxation. This seams reasonable given that the conversion rate would have to increase by many orders of magnitude, relative to typical gas-phase rates [73], for these effects to be important.

The dipole moment of propargyl radical was determined from the Stark-spectrum of the $v_1$ band of propargyl radical, shown in Figure 4,

corresponding to a static electric field of 51.2 kV/cm. The laser polarization was aligned parallel to the static field, yielding $\Delta M=0$ selection rules. The electric fields were calibrated by measuring the splitting of the gas-phase HCN R(1) transition ($\nu_1$ band) using the same electrode configuration. Since the ground and excited state dipole moments are known for this case [74,75], the splitting provides a direct measurement of the electric field strength. The simulation of the Stark-spectrum is performed with an asymmetric rotor Stark program developed in our group [76]. The full $M$-matrices are set up in an asymmetric rotor basis and diagonalized to calculate Stark and pendular state spectra [76]. Again the non-thermal ground-state populations due to nuclear spin-statistics are carefully taken into account. Using the inertial parameters given in Table 3 and a rotational temperature of 0.37 K, we fit the spectrum and determined the ground state dipole moment $\mu_0$ = -0.150 (5) D and the change in the dipole moment upon vibrational excitation $\Delta\mu$ = 0.02 (1) D. Although in this experiment we only obtain the absolute value of the ground state dipole moment, the sign of $\mu_0$ is determined by comparison with *ab initio* calculations [1]. The resulting calculated spectrum is shown in Figure 4.

To our knowledge this is the first experimental measurement of the permanent electric dipole moment of the propargyl radical, although from the power dependence of the signal intensity in the FT microwave spectrum Tanaka et al. concluded that the dipole moment value was <0.3 D [51]. Botschwina et al. [1] have carried out high level *ab initio* calculations from which they recommend an equilibrium dipole moment of $\mu_e$ = -0.14 (3) D in good agreement with the experimental value. However, this is not really a valid comparison, since the experimental value is vibrationally averaged, while the *ab initio* calculations give the equilibrium value. To address this issue we calculated the v=0 and v=1 vibrational wavefunctions using the Numerov-Cooley method [77] based upon a one-dimensional slice through the corresponding potential energy surface [1]. The resulting expectation values for the CH bond length are $<r_0>=r_e+0.014$ Å and $<r_1>=r_e+0.044$ Å ($r_e$ = 1.0626 Å [1]), for the ground and first vibrationally excited state, respectively. The corresponding averages of the electric dipole moment function of Botschwina et al. [1], obtained by correcting the suggested equilibrium value $\mu_e$ = -0.14 D using the MR-ACPF/EV results [78], yield dipole moments of $<\mu>_0$ = -0.134 D and $<\mu>_1$ = -0.123 D, namely a change upon excitation of $\Delta\mu$ = 0.011 D. From a five-dimensional analysis of all totally symmetric vibrations Botschwina [79] obtained a zero-point correction to the dipole moment of 0.016 D, corresponding to $<\mu>_0$ = -0.124 D. Although this calculation still neglects the non-totally symmetric vibrations and might underestimate the vibrational effects on the ground state dipole moment, it should give a good description of the change in the dipole moment upon excitation of $\nu_1$. Indeed these calculations give a change upon vibrational excitation of $\Delta\mu$ = 0.029 D [79]. Although the effects of vibrational averaging make the agreement between the experimental and theoretical ground state dipole moment somewhat worse, the difference is still well within the given theoretical error of ±0.03 D [1]. The change in dipole moment of $\Delta\mu$ = 0.029 D obtained from the *ab initio* calculations is within the experimental error. Overall the agreement between the *ab initio* calculations and experiment is quite satisfactory.

**Summary**

We reported on the development of a novel effusive pyrolysis source for generating radicals for pick-up in superfluid liquid helium droplets. The source is used to generate propargyl radical and a rotationally resolved spectrum of the radical is reported. Analysis of a Stark spectrum of the radical provides the first experimental measurement of its permanent electric dipole moment, namely $\mu_0 = -0.150\,(5)$ D, and the change upon vibrational excitation, $\Delta\mu = 0.02\,(1)$ D.

This study demonstrates the potential of the helium droplet method to study a wide range of radicals and, in the future, radical complexes. Considering the importance of propargyl and similar radicals in combustion and soot-formation and the unique properties of liquid helium droplets [58,59] we are currently extending this work to the study of pre-reactive propargyl radical dimers and propargyl···Br complexes. Given the ability of helium droplets to form metastable species [58,59], there is considerable potential to stabilize highly reactive complexes, including those containing more than one radical.

**Acknowledgments**

We are grateful to P. Chen for several helpful discussions regarding his pyrolysis source design, to D.T. Moore for the implementation of the population analysis into the Stark program, and to P. Botschwina for calculations of vibrational effects on the dipole moment. This work was supported by the Air Force Office of Scientific Research (AFOSR). Partial support is also acknowledged from the National Science Foundation (CHE-99-87740) and the Alexander von Humboldt Foundation (fellowship for J.K.).

**References**
1. P. Botschwina, R. Oswald, J. Flügge, and M. Horn, *Z. Phys. Chem.* **188,** 29 (1995).
2. E. Hirota: *High-Resolution Spectroscopy of Transient Molecules*, Springer-Verlag, Berlin, 1985.
3. C. L. Morter, C. Domingo, S. K. Farhat, E. Cartwright, G. P. Glass, and R. F. Curl, *Chem. Phys. Lett.* **195,** 316 (1992).
4. L. Yuan, J. DeSain, and R. F. Curl, *J. Mol. Spectrosc.* **187,** 102 (1998).
5. D. W. Kohn, H. Clauberg, and P. Chen, *Rev. Sci. Instrum.* **63,** 4003 (1992).
6. M. R. Cameron and S. H. Kable, *Rev. Sci. Instrum.* **67,** 283 (1996).
7. K. D. Setzer, E. H. Fink, A. B. Alekseyev, H. P. Liebermann, and R. J. Buenker, *J. Mol. Spectrosc.* **206,** 181 (2001).
8. M. Kareev, M. Sablier, and T. Fujii, *J. Phys. Chem. A* **104,** 7218 (2000).
9. R. S. Ram, P. F. Bernath, and K. H. Hinkle, *J. Chem. Phys.* **110,** 5557 (1999).
10. V. Aquilanti, D. Ascenzi, E. Braca, D. Cappelletti, G. Liuti, E. Luzzatti, and F. Pirani, *J. Phys. Chem. A* **101,** 6523 (1997).
11. H. Bai and B. S. Ault, *Chem. Phys. Lett.* **188,** 126 (1992).
12. P. Biggs, A. A. Boyd, C. E. Canosamas, D. M. Joseph, and R. P. Wayne, *Measurement Science & Technology* **2,** 675 (1991).
13. J. U. Grabow, N. Heineking, and W. Stahl, *Z. Naturforsch. A* **46,** 914 (1991).
14. F. J. Lovas, R. D. Suenram, T. Ogata, and S. Yamamoto, *Astrophys. J.* **399,** 325 (1992).
15. D. T. Anderson, S. Davis, T. S. Zwier, and D. J. Nesbitt, *Chem. Phys. Lett.* **258,** 207 (1996).
16. S. Davis, M. Farnik, D. Uy, and D. J. Nesbitt, *Chem. Phys. Lett.* **344,** 23 (2001).
17. L. Li, J. T. Graham, and W. Weltner, *J. Phys. Chem. A* **105,** 11018 (2001).
18. H. Wang, Z. Lu, and F. A. Kong, *Chinese Chemical Letters* **12,** 971 (2001).
19. M. L. Tsao, Z. D. Zhu, and M. S. Platz, *J. Phys. Chem. A* **105,** 8413 (2001).
20. K. Kobayashi and T. J. Sears, *Can. J. Phys.* **79,** 347 (2001).
21. C. A. Taatjes and J. F. Hershberger, *Annu. Rev. Phys. Chem.* **52,** 41 (2001).
22. I. U. Goldschleger, A. V. Akimov, E. Y. Misochko, and C. A. Wight, *J. Mol. Spectrosc.* **205,** 269 (2001).
23. G. Bucher, M. Halupka, C. Kolano, O. Schade, and W. Sander, *Eur. J. Org. Chem.* **2001,** 545 (2001).
24. B. C. Chang, M. L. Costen, A. J. Marr, G. Ritchie, G. E. Hall, and T. J. Sears, *J. Mol. Spectrosc.* **202,** 131 (2000).
25. H. J. Deyerl, I. Fischer, and P. Chen, *J. Chem. Phys.* **110,** 1450 (1999).
26. N. M. Marinov, W. J. Pitz, C. K. Westbrook, A. M. Vincitore, M. J. Castaldi, S. M. Senkan, and C. F. Melius, *Combust. Flame* **114,** 192 (1998).
27. P. R. Westmoreland, A. M. Dean, J. B. Howard, and J. P. Longwell, *J. Phys. Chem.* **93,** 8171 (1989).

28. U. Alkemade and K. H. Homann, *Zeitschrift fur Physikalische Chemie Neue Folge* **161,** 19 (1989).
29. S. E. Stein, J. A. Walker, M. M. Suryan, and A. Fahr, *Symp. (Int.) Combust.* **23,** 85 (1990).
30. P. Botschwina, M. Horn, R. Oswald, and S. Schmatz, *J. Electron. Spectrosc.* **108,** 109 (2000).
31. H. Honjou, M. Yoshimine, and J. Pacansky, *J. Phys. Chem.* **91,** 4455 (1987).
32. A. Hinchcliffe, *J. Mol. Struct.* **37,** 295 (1977).
33. C. H. Wu and R. D. Kern, *J. Phys. Chem.* **91,** 6291 (1987).
34. N. M. Marinov, M. J. Castaldi, C. F. Melius, and W. Tsang, *Combust. Sci. Technol.* **128,** 295 (1997).
35. I. Glassman, *Symp. (Int.) Combust.*, **22,** 295**,** (1988).
36. J. A. Miller, *Symp. (Int.) Combust.*, **26,** 461**,** (1996).
37. J. M. Goodings, D. K. Bohme, and C.-W. Ng, *Combust. Flame* **36,** 27 (1979).
38. D. B. Olson and H. F. Calcote, *Symp. (Int.) Combust.* **18,** 453 (1981).
39. R. W. Fessender and R. H. Schuler, *J. Chem. Phys.* **38,** 2147 (1963).
40. P. H. Kasai, *J. Am. Chem. Soc.* **94,** 5950 (1972).
41. D. A. Ramsey and P. Thistlethwaite, *Can. J. Phys.* **44,** 1381 (1966).
42. M. E. Jacox and D. E. Milligan, *Chem. Phys.* **4,** 45 (1974).
43. M. Wyss, E. Riaplov, and J. P. Maier, *J. Chem. Phys.* **114,** 10355 (2001).
44. J. W. Huang and W. R. Graham, *J. Chem. Phys.* **93,** 1583 (1990).
45. G. Maier, H. P. Reisenauer, W. Schwab, P. Carsky, V. Spirko, B. A. Hess, and L. J. Schaad, *J. Chem. Phys.* **91,** 4763 (1989).
46. J. M. Oakes and G. B. Ellison, *J. Am. Chem. Soc.* **105,** 2969 (1983).
47. D. W. Minsek and P. Chen, *J. Phys. Chem.* **94,** 8399 (1990).
48. B. Nagels, P. Bakker, L. J. F. Hermans, and P. L. Chapovsky, *Physical Review A* **57,** 4322 (1998).
49. H. J. Deyerl, I. Fischer, and P. Chen, *J. Chem. Phys.* **111,** 3441 (1999).
50. K. Tanaka, T. Harada, K. Sakaguchi, K. Harada, and T. Tanaka, *J. Chem. Phys.* **103,** 6450 (1995).
51. K. Tanaka, Y. Sumiyoshi, Y. Ohshima, Y. Endo, and K. Kawaguchi, *J. Chem. Phys.* **107,** 2728 (1997).
52. P. Botschwina, M. Horn, J. Fluegge, and S. Seeger, *J. Chem. Soc. Faraday Trans.* **89,** 2219 (1993).
53. A. Scheidemann, B. Schilling, J. P. Toennies, and J. A. Northby, *Physica. B* **165,** 135 (1990).
54. R. Frochtenicht, J. P. Toennies, and A. F. Vilesov, *Chem. Phys. Lett.* **229,** 1 (1994).
55. S. Goyal, D. L. Schutt, and G. Scoles, *J. Phys. Chem.* **97,** 2236 (1993).

56. F. Stienkemeier, W. E. Ernst, J. Higgins, and G. Scoles, *J. Chem. Phys.* **102,** 615 (1995).
57. K. K. Lehmann and G. Scoles, *Science* **279,** 2065 (1998).
58. K. Nauta and R. E. Miller, *Science* **283,** 1895 (1999).
59. K. Nauta and R. E. Miller, *Science* **287,** 293 (2000).
60. K. Nauta and R. E. Miller, *J. Chem. Phys.* **113,** 10158 (2000).
61. K. Nauta and R. E. Miller, *J. Chem. Phys.* **111,** 3426 (1999).
62. J. Küpper, J. M. Merritt, and R. E. Miller, *Free Radical Complexes in Helium Nanodroplets: High Resolution Infrared Spectroscopy*, in 26. International Symposium on Free Radicals, (2001), 19, Assisi, Italy, (2001).
63. S. R. Yates and J. Gan, *J. Agric. Food Chem.* **46,** 755 (1998).
64. Y. R. Lee and S. M. Lin, *J. Chem. Phys.* **108,** 134 (1998).
65. J. C. Evans and R. A. Nyquist, *Spectrochim. Acta* **19,** 1153 (1963).
66. T. Shimanouchi, *J. Phys. Chem. Ref. Data* **6,** 993 (1977).
67. A. K. Wilson, D. E. Woon, K. A. Peterson, and T. H. Dunning, *J. Chem. Phys.* **110,** 7667 (1999).
68. H. J. Werner, P. J. Knowles, J. Almlof, R. D. Amos, A. Berning, M. J. O. Deegan, F. Eckert, S. T. Elbert, C. Hampel, R. Lindh, W. Meyer, A. Nicklass, K. Peterson, R. Pitzer, A. J. Stone, P. R. Taylor, M. E. Mura, P. Pulay, M. Scheutz, H. Stoll, T. Thorsteinsson, and D. L. Cooper: *MOLPRO, a package of ab initio programs,* version 2002.1, Birmingham, UK, 2002.
69. K. K. Lehmann, *Mol. Phys.* **97,** 645 (1999).
70. K. Nauta and R. E. Miller, *J. Chem. Phys.* **113,** 9466 (2000).
71. K. Nauta and R. E. Miller: The Spectroscopy of Molecules and Unique Clusters in Superfluid Liquid Helium Droplets, in *Atomic and Molecular Beams: State of the Art*, edited by R. Campargue (2000).
72. M. Hartmann, R. E. Miller, J. P. Toennies, and A. F. Vilesov, *Phys. Rev. Lett.* **75,** 1566 (1995).
73. P. L. Chapovsky and L. J. F. Hermans, *Annu. Rev. Phys. Chem.* **50,** 315 (2000).
74. T. E. Gough, R. E. Miller, and G. Scoles, *Faraday Disc.* **71,** 77 (1981).
75. W. L. Ebenstein and J. S. Muenter, *J. Chem. Phys.* **80,** 3989 (1984).
76. D. T. Moore, L. Oudejans, and R. E. Miller, *J. Chem. Phys.* **110,** 197 (1999).
77. J. W. Cooley, *Math. Comput.* **15,** 363 (1961).
78. M. Horn, M. Oswald, R. Oswald, and P. Botschwina, *Phys. Chem. Chem. Phys.* **99,** 323 (1995).
79. P. Botschwina, *private communication*. 2002.

**Tables:**

Table 1: Calculated equilibrium structures, rotational constants, and $\nu_1$ harmonic frequency and transition moment orientation for the electronic ground state of propargyl bromide. See text for details.

|  | HF/cc-pVTZ | MP2/cc-pVTZ | CCSD(T)/cc-pVTZ |
|---|---|---|---|
| $r(C_3,H_3)$ (pm) | 105.36 | 106.11 | 106.32 |
| $r(C_1,H_1)$ (pm) | 107.57 | 108.56 | 108.68 |
| $r(C_2,C_3)$ (pm) | 118.05 | 121.44 | 121.08 |
| $r(C_1,C_2)$ (pm) | 145.38 | 144.64 | 145.48 |
| $r(C_1,Br)$ (pm) | 195.38 | 195.25 | 196.50 |
| $\alpha(H_3,C_3,C_2)$ (°) | 180.0 | 179.9 | 179.7 |
| $\alpha(C_1,C_2,C_3)$ (°) | 180.0 | 180.6 | 180.8 |
| $\alpha(C_2,C_1,H_1)$ (°) | 111.4 | 111.6 | 111.5 |
| $\alpha(C_2,C_1,Br)$ (°) | 111.7 | 111.7 | 111.8 |
| $\phi(H_1,C_1,C_2,Br)$ (°) | 118.5 | 118.6 | 118.4 |
| A (cm$^{-1}$) | 0.7082 | 0.7029 | 0.7019 |
| B (cm$^{-1}$) | 0.0724 | 0.0719 | 0.0711 |
| C (cm$^{-1}$) | 0.0665 | 0.0660 | 0.0653 |
| $\nu_1$ (cm$^{-1}$) | 3621 | 3498 | 3457[a] |
| $\theta$ (°) | 41 | 40 | 40[a] |

[a] The frequency calculation was performed at the CCSD(T)/cc-pVDZ level.

Table 2: Observed rovibrational transitions of the $\nu_1$ band of propargyl radical in helium droplets. The frequencies were obtained by simultaneously fitting Lorentzian profiles to the individual peaks in the field-free spectrum. More transitions are evident in the field-free spectrum but are too weak or strongly overlapped for the frequencies to be accurately determined. The estimated uncertainties of the absolute frequencies are 0.01 cm$^{-1}$, relative uncertanties are 0.0002 cm$^{-1}$ for $K_a$=0 and 0.001 cm$^{-1}$ for $K_a$=1 lines.

|  | Transition | experimental frequency (cm$^{-1}$) |
|---|---|---|
| $K_a$=0 | $3_{03} \leftarrow 2_{02}$ | 3322.782 |
|  | $2_{02} \leftarrow 1_{01}$ | 3322.605 |
|  | $1_{01} \leftarrow 0_{00}$ | 3322.388 |
|  | $0_{00} \leftarrow 1_{01}$ | 3321.916 |
|  | $1_{01} \leftarrow 2_{02}$ | 3321.687 |
| $K_a$=1 | $2_{12} \leftarrow 1_{11}$ | 3322.554 |
|  | $1_{10} \leftarrow 1_{11}$ | 3322.182 |
|  | $1_{11} \leftarrow 1_{10}$ | 3322.105 |

Table 3: Summary of the molecular constants for the propargyl radical in superfluid liquid helium droplets, compared with those obtained in gase-phase studies [4,50,51]. Numbers in paranthesis are one estimated standard deviation.

| Constant | Helium droplet | gas-phase[a] |
|---|---|---|
| $A''$ (cm$^{-1}$) | 9.60847[b] | 9.60847 (18) |
| $(B''+C'')/2$ (cm$^{-1}$) | 0.1198 (5) | 0.312386 (12) |
| $B''-C''$ (cm$^{-1}$) | 0.0035 (2) | 0.0105762 (35) |
| $\Delta_N''$ (cm$^{-1}$) | 0.00042 (1) | 7.35 (122) 10$^{-8}$ |
| $A'$ (cm$^{-1}$) | 9.60258[b] | 9.60258 (11) |
| $(B'+C')/2$ (cm$^{-1}$) | 0.1185 (5) | 0.311641 (7) |
| $B'-C'$ (cm$^{-1}$) | 0.0035 (2) | 0.010496 (13) |
| $\Delta_N'$ (cm$^{-1}$) | 0.00062 (1) | 5.37 (76) 10$^{-8}$ |
| $\nu_0$ (cm$^{-1}$) | 3322.15 (1) | 3322.292 (10) |
| $\mu_a''$ (D) | -0.150 (5) | — |
| $\Delta\mu$ (D) | 0.02 (1) | — |

[a] Reported ground state values are from reference [50], excited state values are from reference [4].

[b] $A''$ and $A'$ are fixed at their respective gas-phase values [4,50].

**Figure captions:**

Figure 1: A schematic diagram of the pyrolysis source. The overall length of the glass tube is 72 cm. A copper shield is mounted to a water cooled block, which helps to shield the bolometer from the radiative heat of the heated tip. The helium droplet beam passes approximately 1 mm in front of the heated tube.

Figure 2: a) An experimental spectrum of the $\nu_1$-band of propargyl bromide, embedded in helium droplets. b) A calculated spectrum corresponding to *A'*=*A''*=0.025 cm$^{-1}$, *B'*=*C'*=0.0148 cm$^{-1}$, *B''*=*C''*=0.0152 cm$^{-1}$. See text for details.

Figure 3: a) An experimental infrared spectrum of the $\nu_1$-band of propargyl radical, embedded in helium droplets. b) A simulated spectrum using the molecular constants given in Table 3, a rotational temperature of 0.37 K, and Lorentzian linewidths of 0.0225 cm$^{-1}$ for $K_a$=0 lines and 0.045 cm$^{-1}$ for $K_a$=1 lines.

Figure 4: An experimental Stark-spectrum of the $\nu_1$-band of propargyl radical in helium droplets, at an electric field of 51.2 kV/cm. b) A simulation using the molecular constants given in Table 3, a rotational temperature of 0.37 K, and Lorentzian linewidths of 0.0175 cm$^{-1}$ $K_a$=0 lines and 0.035 cm$^{-1}$ for $K_a$=1 lines.

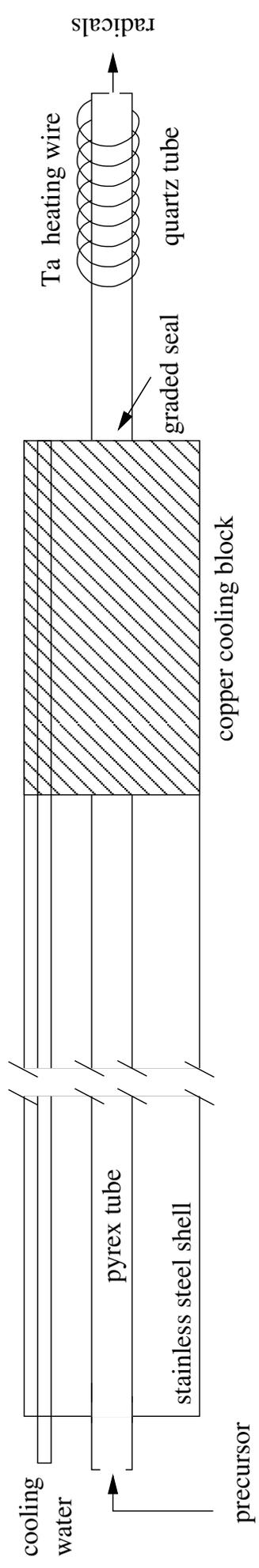

Fig. 1

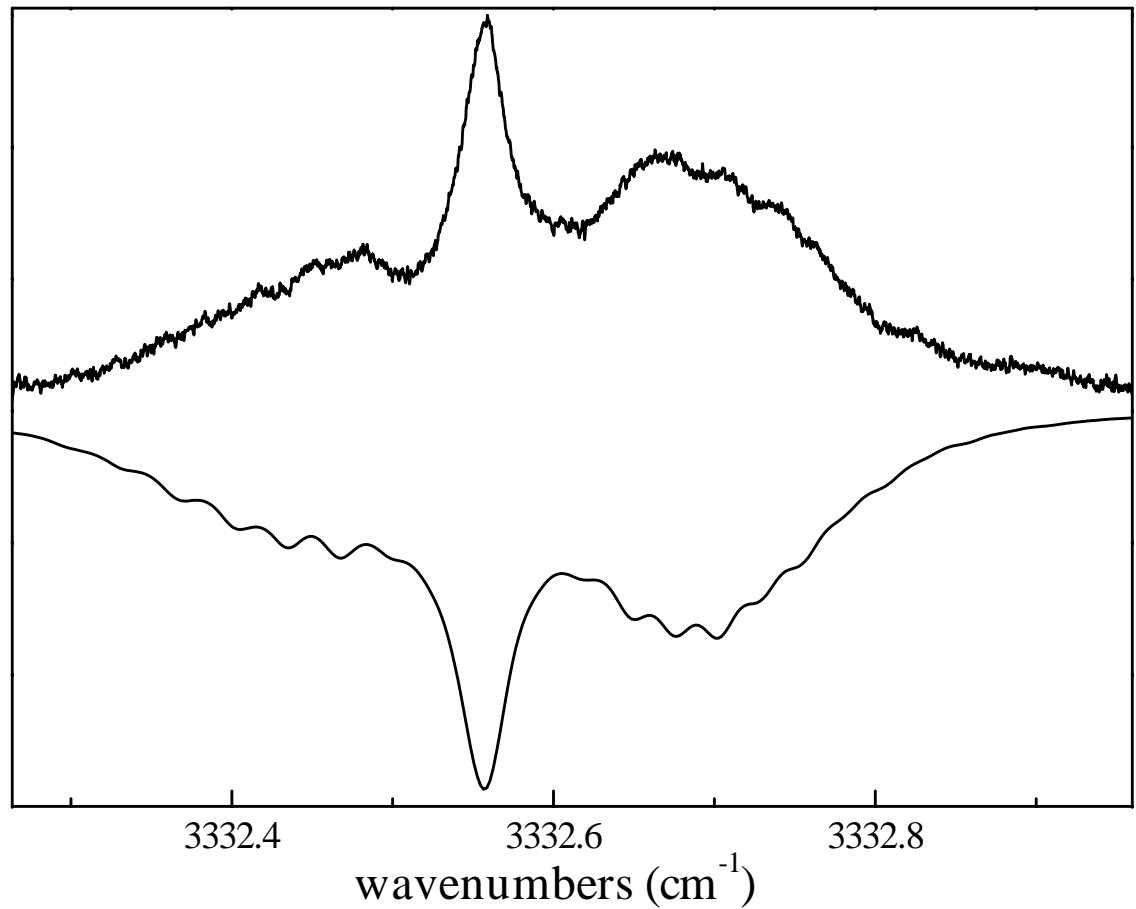

Fig. 2

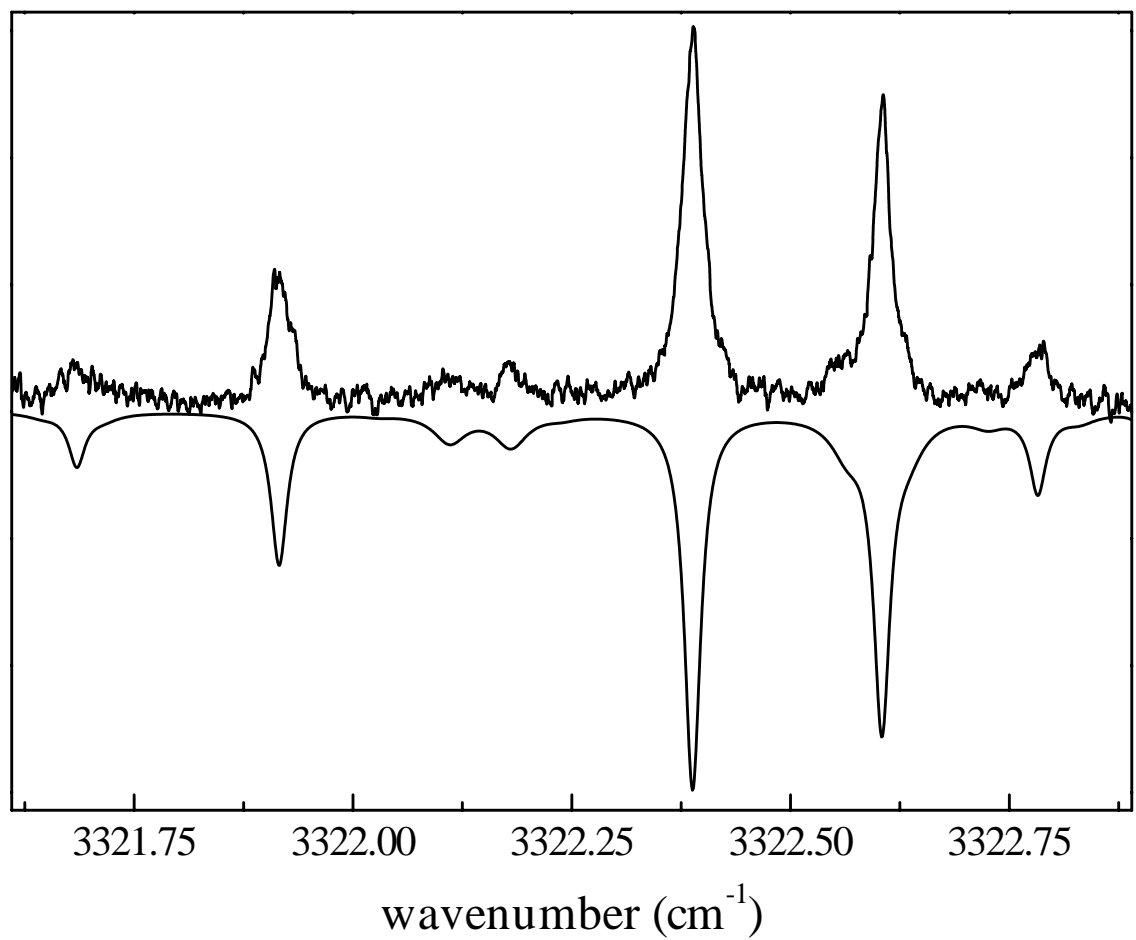

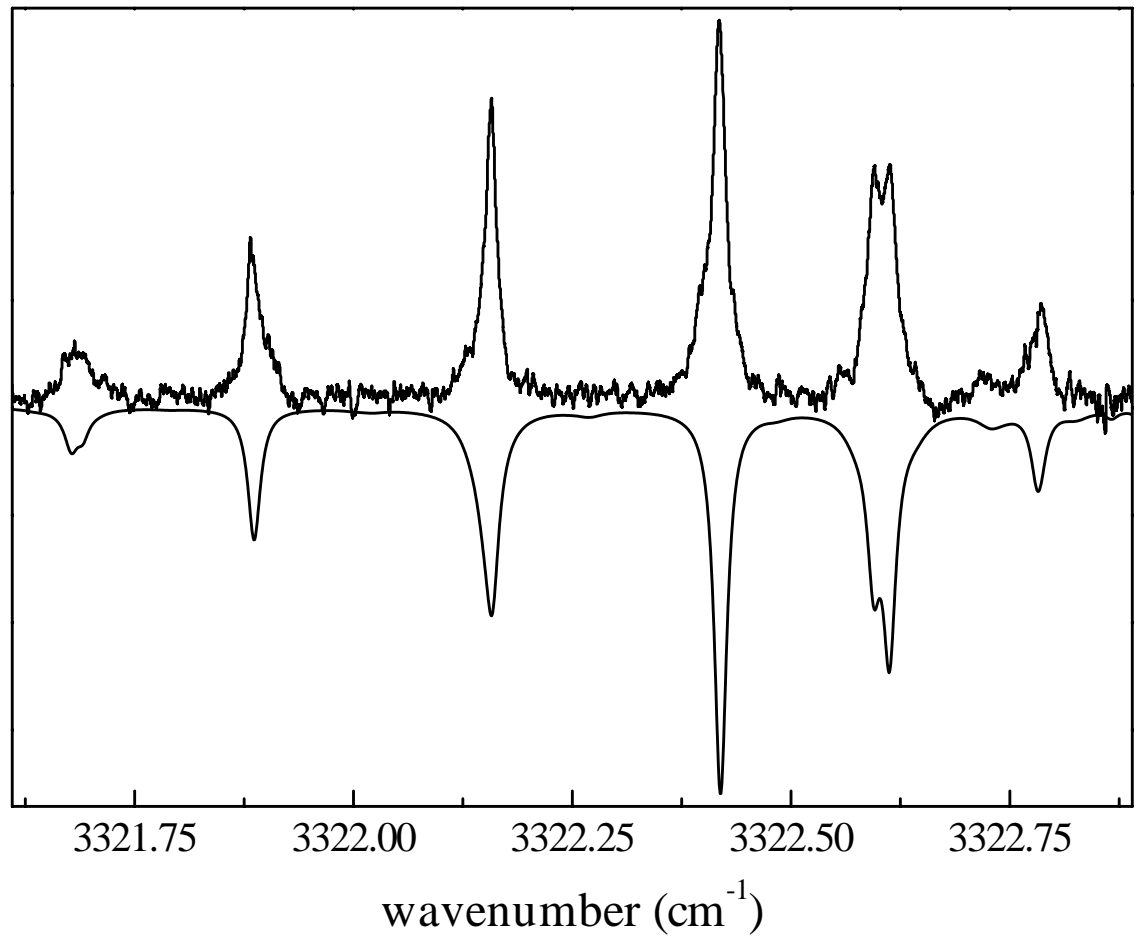

Fig. 4